\documentclass[aps,a4paper,twocolumn]{revtex4}
\usepackage{amsmath}
\usepackage{graphicx}
\usepackage{subfigure}
\usepackage[usenames,dvipsnames]{color}
\definecolor{darkblue}{RGB}{0,0,196}
\usepackage[colorlinks=true,linkcolor=darkblue,citecolor=darkblue,urlcolor=darkblue]{hyperref}
\usepackage{setspace}
\usepackage{url}
\usepackage{hyperref}
\usepackage{ulem}
\usepackage{xcolor}
\hypersetup{
  colorlinks   = true, 
  urlcolor     = red, 
  linkcolor    = blue, 
  citecolor   = blue 
}
\usepackage{graphicx}
\usepackage{amsmath,bbm}
\usepackage{amssymb,bm}
\usepackage{yfonts}
\usepackage{comment}

\usepackage[usenames,dvipsnames]{color}
\definecolor{darkblue}{RGB}{0,0,196}
\usepackage[colorlinks=true,linkcolor=darkblue,citecolor=darkblue,urlcolor=darkblue]{hyperref}
\begin{document}

\title{Rotational effects on the chiral crossover transition in QCD matter}

\author{Nandita Padhan$^{1}$} 
\author{Kshitish Kumar Pradhan$^{2}$}
\author{Arghya Chatterjee$^{1}$\footnote{achatterjee.phy@nitdgp.ac.in}} 
\author{Raghunath Sahoo$^{2}$\footnote{Raghunath.Sahoo@cern.ch}} 
\affiliation{$^{1}$Department of Physics, National Institute of Technology Durgapur, Durgapur 713209, West Bengal, India}
\affiliation{$^{2}$Department of Physics, Indian Institute of Technology Indore, Simrol, Indore 453552, India}

\begin{abstract}
 
We analyze the impact of rotation on the chiral crossover transition of QCD matter within the framework of a hadron resonance gas model. By extending the conventional formulation of the renormalized chiral condensate to a rotating hadronic medium, we find that rotation significantly modifies the chiral condensate and systematically suppresses the pseudocritical temperature as the angular velocity increases. The pseudocritical line in the $T-\omega$ plane is examined, and the rotational dependence of the pseudocritical temperature is quantified through leading- and next-to-leading-order rotational curvature coefficients.
We further investigate the combined effects of angular velocity and baryon chemical potential and find that their interplay leads to a stronger suppression of the pseudocritical temperature. Additionally, the spatial dependence of the chiral crossover transition is examined through the radial variation of the pseudocritical temperature. We find that the pseudocritical temperature decreases with increasing radial distance, indicating that the chiral crossover sets in at progressively lower temperatures as one goes away from the rotation axis.

\end{abstract}
\date{\today}
\maketitle
\section{Introduction}

The strong motivation behind relativistic heavy-ion collision experiments at facilities such as the Large Hadron Collider (LHC) and the Relativistic Heavy Ion Collider (RHIC) is to comprehend the structure of the quantum chromodynamics (QCD) phase diagram. Chiral symmetry breaking, a basic property of the QCD phase diagram, is the cornerstone of our conceptual knowledge of hadron physics \cite{Bochkarev:1984zx, Ko:1991dj, Ko:1991kw}. At low temperature and density, the chiral symmetry is spontaneously broken and is characterized by the non-zero value of chiral condensate
$\langle \bar{\psi}\psi \rangle \neq 0$. At vanishing baryon chemical potential, lattice QCD (lQCD) predicts that this transition is a crossover with a pseudo-critical temperature of $T_{c}=156.5\pm1.5~\mathrm{MeV}$ \cite{HotQCD:2018pds}.
Above this pseudo-critical temperature, the chiral condensate decreases gradually, approaching $\langle \bar{\psi}\psi \rangle \approx 0$, corresponding to the restoration of chiral symmetry. 
The location and nature of the chiral transition are sensitive to the thermodynamic conditions of the system, particularly temperature and baryon density. They can also be influenced by external fields, like magnetic fields and vorticity. Understanding the response of QCD matter under such extreme conditions is therefore essential for a comprehensive characterization of its phase structure.

Recently, a considerable interest has been devoted to investigating the properties of QCD matter, including the deconfinement and chiral phase transitions, in the presence of intense magnetic fields and rotation. 
Strong (but transient) magnetic fields are expected to be produced in non-central heavy-ion collisions ~\cite{Deng:2012pc}, which can notably modify the phase structure of the medium~\cite{Fukushima:2016vix}. 
In addition to the magnetic field, orbital angular momentum carried by the colliding nuclei can induce strong vorticity (rotation) in the produced medium, which can influence the medium properties as well as the phase structure of QCD matter. Various hydrodynamic models have been employed to estimate the magnitude of vorticity. 
In Ref.~\cite{Jiang:2016woz}, the authors have systematically quantified the global angular momentum and vorticity of the quark-gluon plasma. The observation of $\Lambda$ hyperon polarization by the STAR Collaboration~\cite{STAR:2017ckg} provided the first experimental evidence of vorticity of the medium, with an estimated magnitude of $\omega \sim 10^{21}~\mathrm{s}^{-1}$. 
This has stimulated considerable theoretical and experimental interest in investigating the rotational properties of QCD matter. 
In particular, studies based on lQCD calculations have estimated the angular momenta of quarks and gluons in the rotating QCD vacuum~\cite{Yamamoto:2013zwa}. Further insights into the phase structure of QCD matter have been obtained from studies of the possible existence of a critical point associated with the chiral phase transition in the $T$--$\omega$ plane~\cite{Jiang:2016wvv}.
In addition, rotation also affects the thermodynamic~\cite{Fujimoto:2021xix, Pradhan:2023rvf, Mukherjee:2023qvq} and transport properties~\cite{Padhan:2025qhz, Dwibedi:2025boz, Padhan:2024edf}, conserved-charge fluctuations of hadronic matter~\cite{Mukherjee:2023ijv, Sahoo:2025fif}, and meson condensation~\cite{Liu:2017spl, Pradhan:2025pol}. 
Furthermore, in Refs.~\cite{Mukherjee:2023qvq, Padhan:2026mwg}, the authors have demonstrated the effect of rotation on the freeze-out temperature in the $T$ - $\mu_B$ plane. 

A variety of theoretical approaches have been employed to investigate the chiral phase transition and the effects of external magnetic fields and rotation on QCD matter. 
In the presence of a magnetic field, these include magnetic catalysis and inverse magnetic catalysis~\cite{Shovkovy:2012zn, Miransky:2015ava}. Similarly, several QCD-inspired models have predicted that rotation suppresses the chiral transition temperature~\cite{Jiang:2016wvv, Ebihara:2016fwa, Chernodub:2017ref, Wang:2018sur, Sun:2023kuu, Xu:2022hql}.
Holographic studies have also reported a decrease in the critical temperature with increasing angular velocity~\cite{Long:2026rlc, Chen:2022mhf}. In contrast, recent lQCD calculations indicate that both the deconfinement transition temperature and the chiral restoration temperature increase with angular velocity~\cite{Braguta:2022str, Braguta:2021jgn, Yang:2023vsw}, highlighting a discrepancy between lQCD results and the predictions of effective QCD and holographic approaches. This calls for further investigation to obtain a more comprehensive understanding of the effects of rotation on QCD thermodynamics and its phase structure. At temperatures below the pseudo-critical temperature, the thermodynamics of QCD matter can be described to good accuracy within the hadron resonance gas (HRG) framework~\cite{Karsch:2004ti,Ratti:2010kj}. The HRG model has been successfully employed to describe thermodynamic observables and conserved charge fluctuations calculated in lQCD~\cite{Karsch:2003zq, Ejiri:2005wq, Cheng:2008zh}, and provides an effective framework for studying the chemical freeze-out conditions in heavy-ion collisions~\cite{Cleymans:2005xv, Becattini:2005xt, Andronic:2008gu, Lysenko:2024hqp}. In addition, the temperature-dependent chiral condensate has also been investigated within the HRG framework~\cite{Toublan:2004ks, Borsanyi:2010bp, Tawfik:2005qh, Andersen:2022clu}.
For a pion gas, the pseudocritical temperature found in next-to-next-to-leading-order (NNLO) chiral perturbation theory can go as high as $250~\mathrm{MeV}$~\cite{Gerber:1988tt, GarciaMartin:2006jj}, which decreases to $190~\mathrm{MeV}$  when more hadronic states are included. However, since the dependence of hadron masses on the quark masses was not sufficiently well constrained, the uncertainties associated with HRG calculations were difficult to quantify. In a recent study ~\cite{Biswas:2022vat}, the authors revisited the chiral condensate within the HRG model and systematically quantified these uncertainties, demonstrating that improved procedures yield a more precise estimate of $T_c$. Subsequently, the same framework was extended to investigate the effects of repulsive mean-field interactions in the HRG model~\cite{Biswas:2024xxh}. 

Given that the HRG framework provides a well-established description of low-temperature QCD thermodynamics and the temperature dependence of the chiral condensate, it offers a natural approach to studying rotational effects on the chiral crossover. In this work, we systematically investigate, for the first time, the rotational dependence of the chiral crossover transition within the HRG model.
We use the standard formulation of the renormalized chiral condensate within the rotating HRG framework. Previous studies have reported the curvature coefficients of the chiral crossover transition line in the $T$--$\mu_B$ plane within the HRG framework and found them to be consistent with lQCD results~\cite{Biswas:2022vat}. 
We follow a similar parametrization to characterize the rotational dependence of the chiral transition temperature in the $T$ - $\omega$ plane and extract the corresponding rotational curvature coefficients. 
We further examine the dependence of the chiral transition temperature on the combined effect of rotation and baryon chemical potential. The results showing the spatial dependence of the chiral crossover transition are also demonstrated. Despite growing interest in rotating QCD matter, a systematic understanding of the effects of rotation on the chiral transition within the hadronic medium remains unexplored. Our results provide new insights into the chiral transition and deepen our understanding of its dependence on rotation.

The outline of this paper is as follows. In Sec.~\ref{sec_formulation}, we discuss the HRG model and the contributions of light- and heavy-mass hadrons to the renormalized chiral condensate. We then extend the formulation to the rotating HRG framework to investigate the effect of rotation on the chiral crossover transition.
In Sec.~\ref{sec_results}, we present our numerical results. Finally, we summarize our main findings and present our conclusions in Sec.~\ref{sce_summary}.

\section{Formalism}
\label{sec_formulation}
\subsection{Hadron Resonance Gas model}
The HRG model provides an effective description of the thermodynamic properties of strongly interacting matter in the hadronic phase. Hadronic interactions in the HRG model can be effectively incorporated by treating the contributions of narrow resonances as stable hadronic states, an approximation supported by S-matrix calculations~\cite {Dashen:1969ep, Dashen:1974jw}. In the Grand Canonical Ensemble (GCE), the total pressure is obtained by summing the contributions from all hadrons and resonances and can be expressed as~\cite{Pradhan:2023rvf, Andronic:2012ut}
\begin{equation}
\label{eq_normalP}
    P^{id}_i(T,\mu_i) = \pm \frac{Tg_i}{2\pi^2} \int_{0}^{\infty} p^2 dp\ \ln\{1\pm \exp[-(E_i-\mu_i)/T]\}.
\end{equation}
Here, $g_i$ and $E_i=\sqrt{p^2+m_i^2}$ denote the degeneracy and energy of the $i$th hadron, respectively. The $\pm$ signs correspond to fermions and bosons, respectively. Here, $\mu_i$ denotes the chemical potential of the $i$th hadron, given by $\mu_i=B_i\mu_B+S_i\mu_S+Q_i\mu_Q$, where $B_i$, $S_i$, and $Q_i$ are the baryon number, strangeness, and electric charge quantum numbers, respectively, while $\mu_B$, $\mu_S$, and $\mu_Q$ represent the corresponding chemical potentials associated with baryon number, strangeness, and electric charge.

\subsection{Renormalized Observables for Chiral Condensation}
The light quark mass derivative of the pressure can be used to calculate the light quark condensate (also known as chiral condensate) at finite temperature, which is expressed as
\begin{equation}
\langle\bar{\psi}\psi\rangle_{l,T}=\langle\bar{\psi}\psi\rangle_{l,0} + 
\frac{\partial P}{\partial m_l}~,
\label{eq.defppbar}
\end{equation} 

where $m_l$ denotes the light quark mass. In the present work, we consider two degenerate light quark flavors, such that $m_u=m_d=m_l$. At zero temperature, the light quark condensate, $\langle\bar{\psi}\psi\rangle_{l,0}$, can be obtained from the derivative of the vacuum pressure with respect to $m_l$. However, for nonzero light quark masses ($m_l\neq0$), the condensate contains both multiplicative and additive ultraviolet divergences. To eliminate these divergences and construct a finite quantity that is invariant under renormalization-group transformations, we consider the following renormalized combination~\cite{Biswas:2022vat}
\begin{eqnarray}
\label{condensate}
-m_s\left[\langle\bar{\psi}\psi\rangle_{l,T}
-\langle\bar{\psi}\psi\rangle_{l,0}\right]=-m_s\frac{\partial P}{\partial m_l}~,
\end{eqnarray}
where $m_s$ denotes the strange quark mass. With an alternative normalization factor, the renormalized chiral condensate can be expressed as~\cite{Borsanyi:2010bp}
\begin{equation}
 \langle\bar{\psi}\psi\rangle_{R} = 
-\frac{m_l}{m_\pi^4} \left[\langle\bar{\psi}\psi\rangle_{l,T}-
\langle\bar{\psi}\psi\rangle_{l,0}\right]~.
\label{Eq.psibarpsi_R}   
\end{equation}

Another useful indicator of chiral symmetry breaking at finite quark masses is provided by the dimensionless chiral condensate, defined as~\cite{Bazavov:2011nk}
\begin{eqnarray}
\Delta^l_R=d+ m_s r_1^4 \left[\langle\bar{\psi}\psi\rangle_{l,T}-
\langle\bar{\psi}\psi\rangle_{l,0}\right]~,
\label{Eq.relation1}
\end{eqnarray}
where $r_1$ is determined from the static quark potential~\cite{Aubin:2004wf}, while $d=r_1^4m_s(\lim_{m_l\to0}\langle\bar{\psi}\psi\rangle_{l,0})^R$, with the superscript $R$ denoting the renormalized quantity. This definition is based on the property that the light-quark condensate undergoes only multiplicative renormalization in the chiral limit. Considering $(\lim_{m_l\to0}\langle\bar{\psi}\psi\rangle_{l,0})^R=2\Sigma$, together with $\Sigma^{1/3}=272\pm5$ MeV, $m_s=92.2\pm1.0$ MeV in the $\overline{\rm MS}$ scheme at $\mu=2$ GeV for the $2+1$ flavor case~\cite{FlavourLatticeAveragingGroupFLAG:2021npn}, and $r_1=0.3106$ fm~\cite{MILC:2010hzw}, we use $d=0.022791$, as reported in Ref.~\cite{Biswas:2022vat}.

The quark-mass dependence of hadron masses plays a crucial role in determining the chiral condensate. To evaluate the contributions of different hadronic states, including resonances, we require their dependence on the light-quark mass, $m_l$. 
We treat the pseudo-Goldstone bosons, namely pions and kaons, separately from the heavier hadrons and resonances. The contributions of pions and kaons to the chiral condensate, $m_s\,(\partial P/\partial m_l)$, can be written as
\begin{align}
m_s \frac{\partial P}{\partial m_l}=-\frac{m_s}{m_l} & \sum_{i=\pi,K} \frac{g_{i}}{2 \pi^2} \int_0^{\infty} dp~ p^2 \nonumber \\ 
&\times\frac{1}{\exp[({E_i-\mu_i)/T]-1}} ~ \frac{1}{2 E_i} m_l \frac{\partial M_{i}^2}{\partial m_l}.
\label{Eq:PSppbar}
\end{align}
In the present study, we adopt the SU(2) $\chi$PT result for the pion mass~\cite{Gasser:1983yg} and its extended formulation for the kaon mass~\cite{RBC-UKQCD:2008mhs}.
The relevant constants governing the dependence of the pion mass on the light quark mass are taken from Ref.~\cite{Biswas:2022vat}. 
Near the physical point, lQCD results are well described by the relation $m_l,\partial M_\pi/\partial m_l \simeq M_\pi/2$~\cite{Bali:2016lvx}, which provides the essential input for evaluating the pion contribution to the chiral condensate.

The contribution of heavier hadrons and resonances to the chiral condensate can be expressed as
\begin{align}
\label{eq_k1}
m_s \frac{ \partial P}{\partial m_l} =-\frac{m_s}{m_l}& \sum_{i}
\frac{ g_{i}}{2 \pi^2}\int_0^{\infty} dp~ p^2 \nonumber\\
&\times\frac{1}{\exp[({E_i-\mu_i)/T]\pm1}}\frac{M_{i}}{E_{i}} \sigma_{i},
\end{align}
where we use $\sigma_{i} = m_l\frac{\partial M_{i}}{\partial m_l}$. It is to be noted that the $\sigma$ terms can be defined as~\cite{Biswas:2022vat}
\begin{equation}
\begin{aligned}
\sigma_{i} &= m_l
\left.
\frac{\partial M_{i}}
{\partial m_l}
\right|_{m_l=m_l^{\rm phys}}\nonumber \\
&=m_l \langle i|
\bar{u}u+\bar{d}d|i\rangle = M_{\pi}^{2} \left.\frac{\partial M_{i}} {\partial M_{\pi}^{2}}
\right|_{M_{\pi}=M_{\pi}^{\rm phys}},
\end{aligned}
\label{Eq:SigmaDef}
\end{equation}
where $m_l^{\mathrm{phys}}$ and $M_\pi^{\mathrm{phys}}$ represent the physical light-quark mass and pion mass, respectively.
For the ground-state baryons, the $\sigma$ terms are obtained from fits to the masses from low-energy data using chiral effective theory and lQCD~\cite{Copeland:2021qni}. For the remaining hadrons, the lQCD results for the dependence of their masses on the pion mass are considered. 
The corresponding values are listed in Tables I and III of Ref.~\cite{Biswas:2022vat}, respectively.

\subsection{Effects of Rotation on the Chiral Condensate}
We now consider a rotating hadronic medium and formulate the corresponding thermodynamic pressure in the rotating frame. This will allow us to investigate the influence of rotation on the chiral crossover transition. 
For a system rotating at angular velocity $\omega$ about the 
$z$ axis, the rotational coupling to the total angular momentum of the hadronic state causes a shift in the single-particle energy spectrum.
The corresponding dispersion relation can be written as $\varepsilon_l=E-(l+s)\omega$,
where $l$ and $s$ denote the orbital angular momentum quantum number about the rotation axis and the $z$-component of the spin, respectively. 
Consequently, the pressure contribution from the $i$th hadronic species in the rotating medium takes the form~\cite{Fujimoto:2021xix, Mukherjee:2023qvq, Pradhan:2025pol, Padhan:2026mwg}
\begin{align}
\label{eq_pressure}
    P_i &= \pm\frac{T}{8\pi^2} \sum_{\ell=-\infty}^\infty \int dp_r^2 \int  dp_z\; \sum_{\nu = \ell}^{\ell + 2S_i}  J_\nu^2(p_r r) \notag\\
    &\qquad\qquad \times \log\left\{1\pm\exp[-(\varepsilon_{\ell,i} - \mu_i) / T]\right\},
\end{align}
where the single-particle energy in the rotating frame is given by
$\varepsilon_{l,i}=\sqrt{p_r^2+p_z^2+m_i^2}-(l+s)\omega$.
Here, $r$ denotes the radial distance from the rotation axis, and $J_\nu$ is the Bessel function of the first kind. 
The transverse and longitudinal components of the momentum are denoted by the symbols $p_r$ and $p_z$, respectively. 
The rotating system is assumed to be confined within a cylindrical boundary of radius $r=R$, satisfying $R\omega\leq1$ ($c=1$ in natural units). Throughout this work, we fix $R=30~\mathrm{GeV}^{-1}\simeq6$ fm. 
The boundary condition leads to the quantization of the transverse momentum,
$p_r=\xi_{l,i}/R$~\cite{Fujimoto:2021xix},
where $\xi_{l,i}$ denotes the $i$th zero of the Bessel function satisfying $J_l(\xi_{l,i})=0$. Consequently, the transverse momentum becomes discretized, with the effect being particularly important in the low-momentum region. Thus, the lower limit of the $p_r$ integration is effectively shifted from zero to the lowest allowed value, $\xi_{l,i}/R$.

\begin{figure*}[htp!]
\begin{center}
\includegraphics[scale = 0.49]{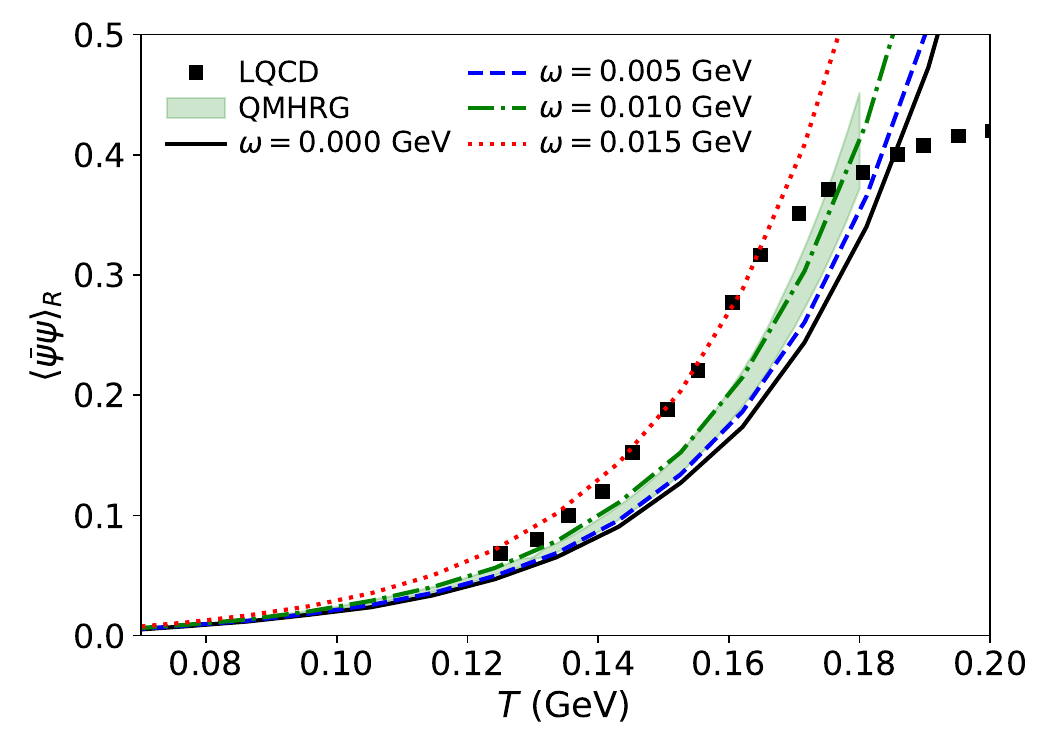}
\includegraphics[scale = 0.5
]{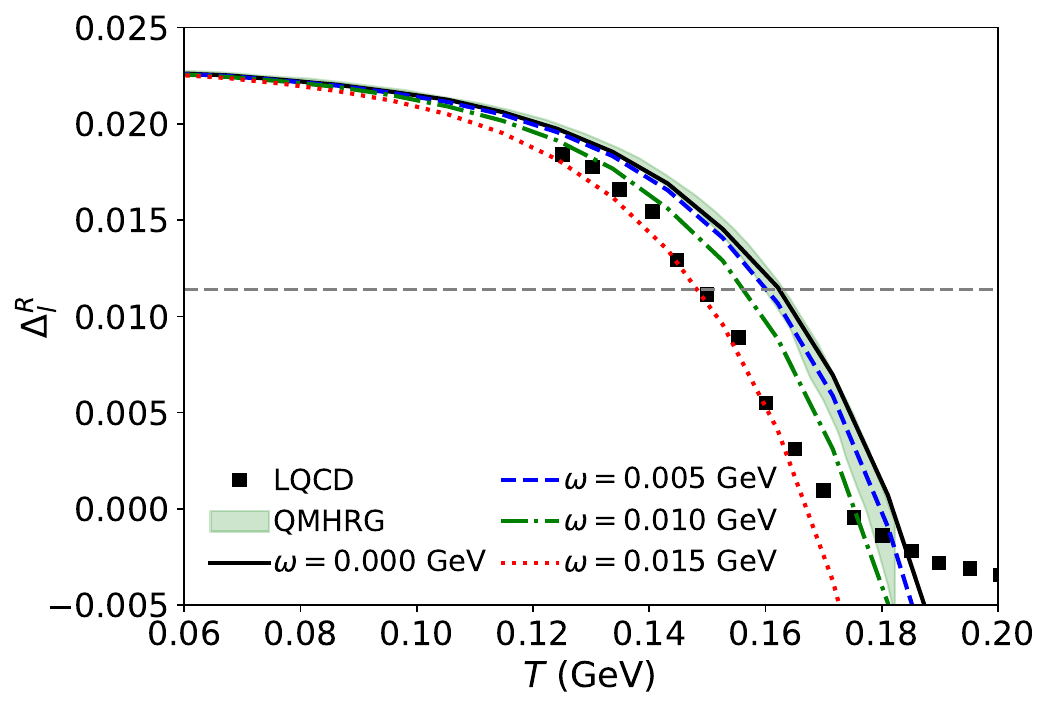}
\caption{(Colour Online) Temperature dependence of the renormalised chiral condensate $\langle \bar{\psi}\psi \rangle_R$ (left) and $\Delta_l^R$ (right). In the non-rotating limit ($\omega=0$), our results are compared with the lQCD results~\cite{Borsanyi:2010bp}, shown by black squares, and the QMHRG results from Ref.~\cite{Biswas:2022vat}, represented by the colored band. Results for finite rotation are shown for $\omega=0.005$, $0.01$, and $0.015~\mathrm{GeV}$.
}
\label{fig1}
\end{center}
\end{figure*}

After determining the pressure from Eq.~(\ref{eq_pressure}), we evaluate its derivative with respect to the light quark mass in the rotating frame. This derivative provides the basis for calculating the chiral condensate in the rotating HRG framework. For example, Eq.~(\ref{eq_k1}) in a rotating hadronic medium is now given by
\begin{align}
\label{rotating_condensate}
m_s \frac{\partial P}{\partial m_l}
=-\frac{m_s}{m_l} &
\sum_i\frac{1}{8\pi^2}
\sum_{\ell=-\infty}^{\infty}
\int dp_r^2\int dp_z
\sum_{\nu=\ell}^{\ell+2S_i}
J_\nu^2(p_r r)
\nonumber\\
&\quad\times \frac{1}{\exp{[(\varepsilon_{l,i}}-\mu_i)/T]\pm 1}
\frac{M_i}{E_i}
\sigma_i .
\end{align}
Here, all quantities retain the definitions given in Eq.~(\ref{condensate}), with the pressure evaluated in the rotating frame. The chiral condensate in the presence of rotation is then obtained from Eq.~(\ref{rotating_condensate}).

\section{Results and Discussion}
\label{sec_results}

We explore the effect of rotation on the chiral crossover transition within the rotating hadron resonance gas framework. All non-interacting hadrons and resonances with masses up to $2.6~\mathrm{GeV}$ are included, as listed by the Particle Data Group~\cite{ParticleDataGroup:2016lqr}. The left panel of Fig.~\ref{fig1} shows the temperature dependence of the renormalized chiral condensate, $\langle\bar{\psi}\psi\rangle_R$, calculated using Eq.~(\ref{Eq.psibarpsi_R}). The solid black line represents the renormalized chiral condensate (in the non-rotating limit) at $\omega = 0$, and is found to increase with temperature. We compare our non-rotating HRG results with the lQCD data~\cite{Borsanyi:2010bp} for the renormalized chiral condensate. The lattice results considered here correspond to $2+1$-flavor QCD with staggered fermions, employing one-link stout improvement and physical quark masses, with $m_s/m_l=28.15$. The same quark-mass ratio is used in our calculations. One can see from the figure that the curve from the lQCD calculation increases monotonically with temperature and exhibits a saturation-like behavior beyond $T\simeq 200~\mathrm{MeV}$. At the chiral crossover temperature $T_c$, no discontinuity is expected in the renormalized chiral condensate, as physical quark masses lead to a crossover rather than a true phase transition. In contrast, our non-rotating HRG results do not exhibit a saturation behavior at higher temperatures and remain systematically below the lQCD results. 
In this context, earlier studies within QMHRG~\cite{Biswas:2022vat, Biswas:2024xxh} have also found a similar behavior for $\langle\bar{\psi}\psi\rangle_R$, where it increases monotonically with temperature. Our results in the non-rotating limit are in close agreement with the QMHRG results ~\cite{Biswas:2022vat}. The slight difference at high temperature can be attributed to the use of different cut-offs in the hadron mass spectrum. In QMHRG, all the hadrons and resonances available from PDG 2016 data Ref.~\cite{ParticleDataGroup:2016lqr} are taken along with those predicted from the quark models~\cite{Capstick:1986ter,Ebert:2009ub}.
Having established this consistency in the non-rotating limit, we extend our analysis to finite rotation by considering angular velocities of $\omega=0.005$, $0.01$, and $0.015$~GeV. The corresponding results are represented by the blue dashed, green dash-dotted, and red dotted lines, respectively. We observe that the same qualitative temperature dependence persists in the presence of rotation. However, at a fixed temperature, the magnitude of the renormalized chiral condensate increases with increasing $\omega$. In particular, the condensate is enhanced by approximately $7\%$ at $\omega=0.005$~GeV and by about $62\%$ at $\omega=0.015$~GeV relative to the non-rotating case at a temperature $T$ = 200 MeV. This pronounced enhancement demonstrates that rotation has a significant effect on the renormalized chiral condensate, with the effect becoming increasingly stronger with increasing magnitude of angular velocity.

To further assess the consistency of the HRG prediction, the renormalised chiral condensate is evaluated using an alternative definition, $\Delta_l^R$, which is related to the previously defined chiral condensate through $\Delta_l^R = d-(m_s/m_l)(r_1m_{\pi})^4\langle\bar{\psi}\psi\rangle_R$. Moreover, this definition is subsequently used to extract the pseudocritical temperature of the chiral crossover transition. The temperature dependence of $\Delta_l^R$ for different magnitudes of rotation is presented in the right panel of Fig.~\ref{fig1}. In the non-rotating limit, represented by the black solid line, our results are consistent with the QMHRG results reported in Ref.~\cite{Biswas:2022vat}. At temperatures below $100~\mathrm{MeV}$, $\Delta_l^R$ remains nearly constant, with its magnitude gradually decreasing as the temperature increases. A similar qualitative behavior is observed in the presence of rotation. The results for $\omega=0.005$, $0.010$, and $0.015$~GeV are represented by the blue dashed, green dash-dotted, and red dotted lines, respectively. 
At low temperatures, $T\lesssim100$ MeV, the effect of rotation on $\Delta_l^R$ remains weak due to the suppressed thermal population of heavier hadronic states. With increasing temperature, higher mass and higher spin states become increasingly populated. Moreover, the factor $e^{(\ell+s)\omega/T}$, arising from the rotational shift of the single-particle energy, leads to a stronger rotational contribution. As a result, the suppression of $\Delta_l^R$ becomes progressively stronger with increasing angular velocity at higher temperatures.

To estimate the pseudocritical temperature for a crossover transition, we must choose an appropriate observable. For a crossover transition, different observables may not yield the same pseudocritical temperature. Unlike a first-order phase transition, where characteristic signatures such as a discontinuity in the density or a divergence of the specific heat occur at a well-defined critical temperature, a crossover is characterized by smooth but rapid changes. Thus, the inflection point of a rapidly varying observable (such as density) or the peak of the specific heat can be used to define the corresponding pseudocritical temperature. These pseudocritical temperatures may differ among observables, but generally lie within the same broad crossover region~\cite{Aoki:2006we}. 
Furthermore, the chiral condensate serves as an order parameter for a genuine chiral phase transition. Although finite quark masses turn the transition into a crossover, the chiral condensate can still be used to identify the corresponding pseudocritical temperature. In lQCD calculations~\cite{Borsanyi:2010bp}, several observables have been employed to determine the pseudocritical temperature, with the inflection point of the renormalized chiral condensate being one commonly used criterion. Using this criterion, the pseudocritical temperature in Ref.~\cite{Borsanyi:2010bp} is estimated to be $T_c$ = 157 $\pm$ 3 MeV. In this work, we use the observable $\Delta_l^R$ to determine the corresponding pseudocritical temperature, allowing for a consistency check with previous QMHRG~\cite{Biswas:2022vat} and lQCD~\cite{Borsanyi:2010bp} studies. However, as seen from the right panel of Fig.~\ref{fig1}, there is no inflection point in the renormalised chiral susceptibility within the HRG model. As a result, it is difficult to directly predict the pseudocritical temperature within this model. We expect the renormalised chiral condensate to gradually decrease from its vacuum value as the temperature rises and the hadronic phase melts to a new state~\cite{Borsanyi:2010bp, Biswas:2022vat}. But the exact condition (the exact condensate value) for this transition is unknown. However, lQCD calculations show that $\Delta_l^R$ decreases to approximately half of its vacuum value near the pseudocritical temperature~\cite{Aubin:2004wf, Bazavov:2013yv}. Motivated by this observation, the authors in Ref.~\cite{Biswas:2022vat} define $T_c$ within the HRG model as the temperature at which the condensate reaches half of its vacuum value, yielding an estimate in better agreement with lQCD results~\cite{Borsanyi:2010bp}. In the present work, we adopt the same criterion to determine $T_c$ within the HRG framework. This provides a consistent prescription for estimating the pseudocritical temperature and allows us to investigate its dependence on rotation. As shown in the right panel of Fig.~\ref{fig1}, the dashed black line is given as a reference to show half of the maximum $\Delta_l^R$. For the non-rotating limit ($\omega$ = 0), we obtain the pseudocritical temperature of $T_c$ = 162.45 MeV, consistent with the previous estimate of 161.2 $\pm$ 1.7 MeV in QMHRG~\cite{Biswas:2022vat}.

\begin{figure}[htp!]%
\begin{flushleft}
\includegraphics[scale=0.48]{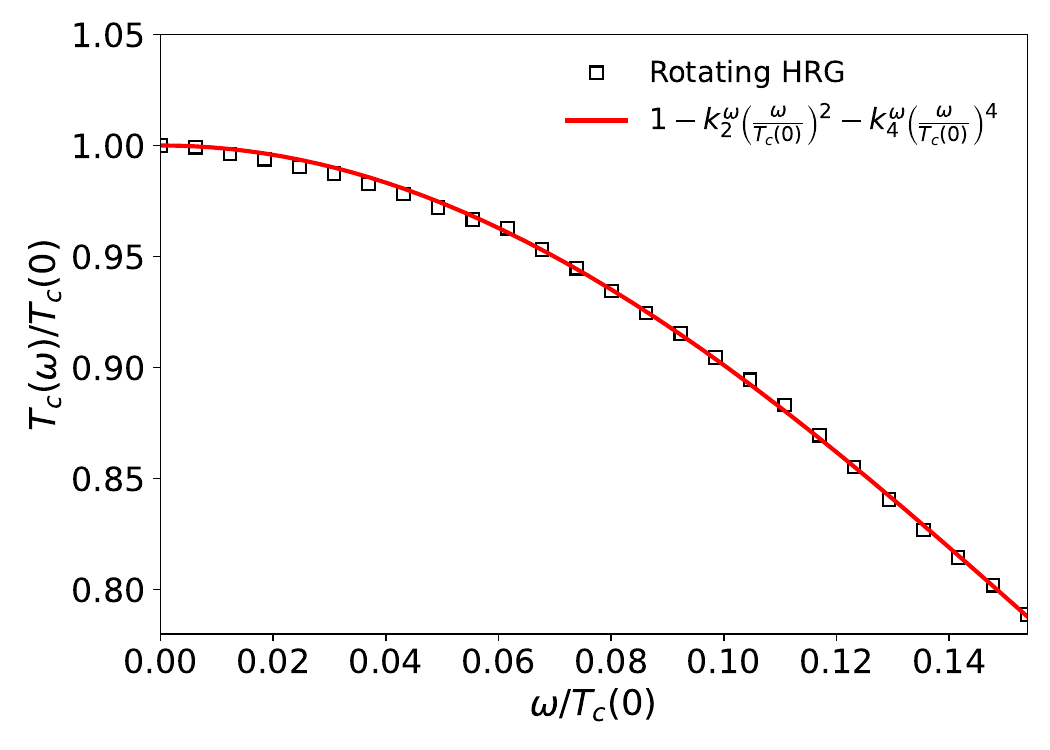}
\end{flushleft}
\caption{(Colour online) The pseudocritical line as a function of $\omega$ scaled to $T_c$ (0). The curvature coefficients extracted from the fit are $k_2^\omega=10.578\pm0.118$ and $k_4^\omega=-68.145\pm6.416$.}
\label{figure2}
\end{figure}

To investigate the effect of rotation on the pseudocritical temperature, we first determine $T_c$ from the $\Delta_l^R$ curves shown in Fig.~\ref{fig1} for finite values of $\omega$. At very low temperatures, the maximum value of $\Delta_l^R$ remains essentially unchanged for the different values of $\omega$ considered and coincides with that in the non-rotating limit. Thus, a common reference value can be used to define the half-maximum criterion for all values of $\omega$. We observe that at higher temperatures, the half of the maximum value of $\Delta_l^R$ happens at progressively lower temperatures with increasing $\omega$, indicating that rotation suppresses the pseudocritical temperature. This behavior is consistent with previous studies of the chiral transition using the NJL~\cite{Chernodub:2017ref, Wang:2018sur}, PNJL~\cite{Sun:2023kuu}, and holographic~\cite{Long:2026rlc, Chen:2022mhf} approaches. 
Similar suppression for the deconfinement transition temperature by rotation has also been reported in several studies~\cite{Braga:2025eiz, Chen:2020ath}, including within the HRG framework~\cite{Mukherjee:2023qvq, Padhan:2026mwg}. 
To further investigate the chiral crossover line as a function of $\omega$ within the HRG framework, Fig.~\ref{figure2} shows the pseudocritical temperature as a function of $\omega$, both normalised to the critical temperature in the non-rotating limit, $T_C$ (0). The half-value criterion based on $\Delta_l^R$, as employed in Refs.~\cite{Aubin:2004wf, Bazavov:2013yv}, is used to find $T_c$ in the presence of rotation. The extracted pseudocritical temperature, shown in open box markers, is found to decrease with increasing angular velocity for the considered range of $\omega$ taken in this study. To quantify this rotational dependence, we follow an approach analogous to that used to characterize the baryochemical-potential dependence of the pseudocritical temperature in lQCD. In lQCD studies, the $\mu_B$ dependence is commonly parametrized through a Taylor expansion around $\mu_B=0$, with the corresponding Taylor coefficients determined from the calculations~\cite{HotQCD:2018pds, Borsanyi:2020fev}. Following this approach, we adopt a similar parametrization to quantify the rotational dependence of the pseudocritical temperature, which can be expressed as

\begin{equation}
\frac{T_c(\omega)}{T_c(0)} = 1 - k_2^\omega \left(\frac{\omega}{T_c(0)}\right)^2 - k_4^\omega \left(\frac{\omega}{T_c(0)}\right)^4 + \mathcal{O}(\omega^6),
\label{eq:Tc_omega_expansion}
\end{equation}
where $k_2^\omega$ and $k_4^\omega$ denote the leading- and next-to-leading-order rotational curvature coefficients, respectively, characterizing the dependence of the pseudocritical temperature on the angular velocity. 
In order to do this, we first obtain the pseudocritical temperature as a function of $\omega$. After normalising by its non-rotating value $T_c(0)$, we fit the ansatz in Eq.~(\ref{eq:Tc_omega_expansion}) using a least-squares fitting method, considering $k_2^\omega$ and $k_4^\omega$ as free parameters. The red solid curve in Fig.~\ref{figure2} represents the fit to the curve obtained for the scaled pseudocritical temperature. The resulting fit yields the curvature coefficients as $k_2^\omega=10.578\pm0.118$ and $k_4^\omega=-68.145\pm6.416$.

\begin{figure}[htp!]%
    \centering
    \begin{flushleft}
    \includegraphics[scale = 0.48]{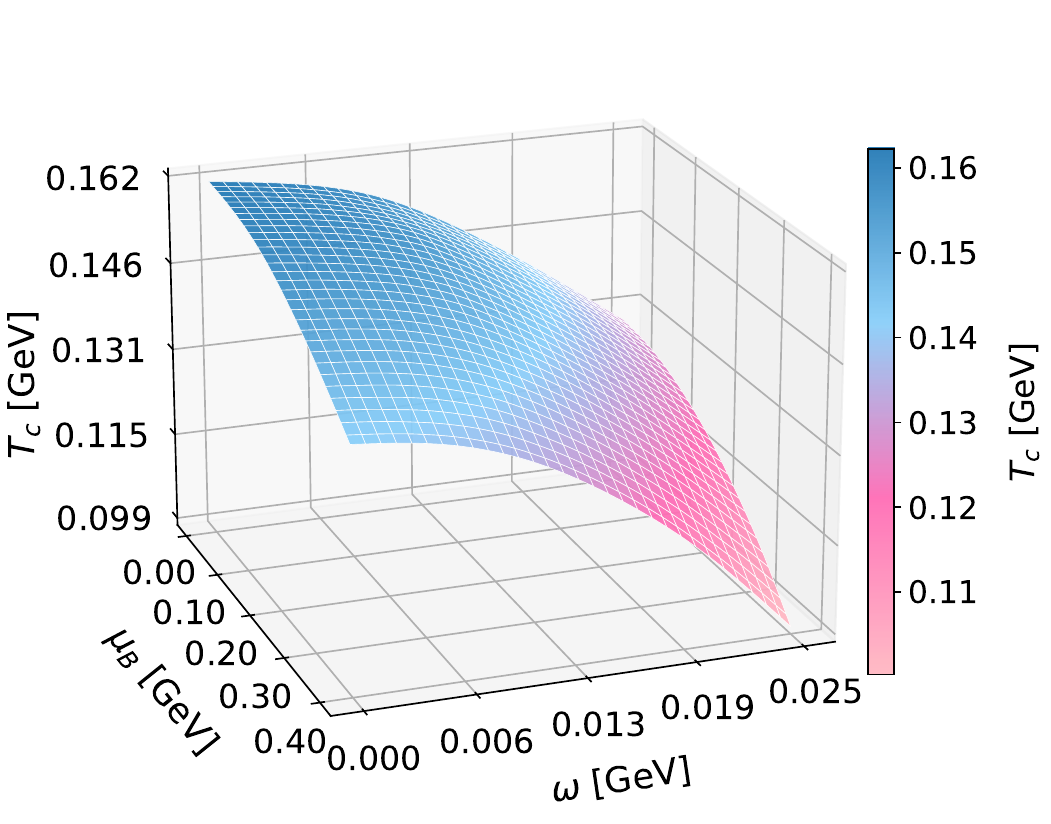}
    \end{flushleft}
    \caption{(Colour online) The pseudocritical temperature as a function of $\mu_B$ and $\omega$.}
    \label{figure3}
\end{figure}
Having established the rotational dependence of the pseudocritical temperature, we now widen our analysis to study its dependence on the combined effect of rotation and baryon chemical potential. This is particularly relevant for low-energy non-central heavy-ion collisions, where both finite baryon density and substantial vorticity are expected to be present. It is observed that the contribution of baryons to the renormalized chiral condensate is smaller than that of mesons and resonances, and is particularly suppressed compared with the pseudoscalar mesons~\cite{Biswas:2022vat}. With increasing $\mu_B$, the baryonic contribution leads to an increasingly pronounced suppression of the chiral condensate, with the effect becoming evident even at relatively low temperatures. This leads to a lower pseudocritical temperature. Further, to demonstrate the interplay between these two effects, the pseudocritical temperature is displayed as a function of $\mu_B$ and $\omega$ in Figure~\ref{figure3}. As expected, the pseudocritical temperature decreases with an increase in both baryon chemical potential and angular velocity. 
However, the combined effects of $\mu_B$ and $\omega$ lead to a stronger suppression of the pseudocritical temperature than that induced by either parameter individually.

\begin{figure}[htp!]%
    \centering
    \begin{flushleft}
    \includegraphics[scale = 0.48]{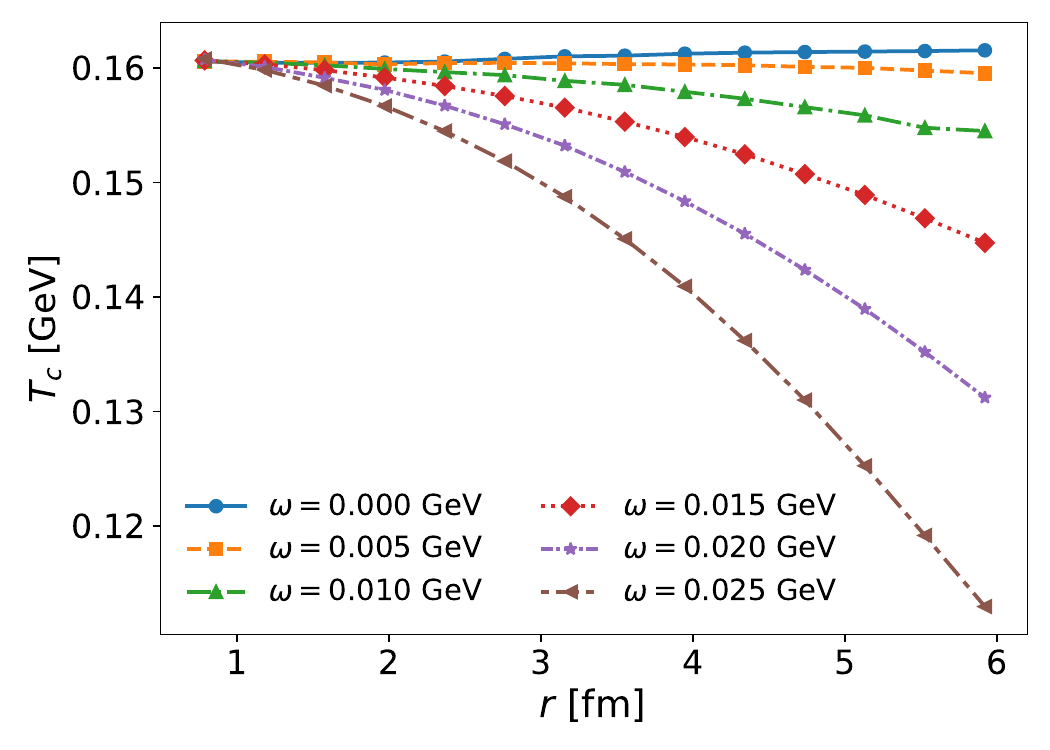}
    \end{flushleft}
    \caption{(Colour Online) Radial dependence of the pseudocritical temperature for $\omega$ in the range 0.005 - 0.025$~\mathrm{GeV}$.}
    \label{fig4}
\end{figure}

This study demonstrates that rotation can significantly influence the phase structure of QCD matter and should therefore be considered when exploring the QCD phase diagram in the $T$--$\mu_B$--$\omega$ parameter space. Moreover, rotation can introduce a spatial dependence in the local thermodynamic properties. It is therefore important to examine whether the pseudocritical temperature exhibits a similar spatial variation. To this end, we investigate the radial dependence of the chiral crossover transition through the variation of the pseudocritical temperature with $r$. In Fig.~\ref{fig4}, we study the radial dependence of the pseudocritical temperature, $T_c$, over the range $1\leq r\leq6$~fm for angular velocities in the range $\omega=0-0.025~\mathrm{GeV}$. We find that the pseudocritical temperature decreases with increasing radial distance, consistent with the corresponding behavior reported in the holographic approach~\cite{Long:2026rlc}. The suppression becomes increasingly pronounced with increasing angular velocity.


\section{Summary}
\label{sce_summary}

In the present work, we investigate the impact of rotation on the chiral crossover transition in ultra-relativistic heavy-ion collisions within the hadron resonance gas (HRG) framework. We extend the standard HRG treatment of the chiral condensate to a rotating hadronic medium by incorporating the coupling of the total angular momentum of the hadronic states to the angular velocity. The renormalized chiral condensate and the corresponding observable $\Delta_l^R$ are employed to characterize the chiral crossover and determine the pseudocritical temperature. Within this framework, we investigate the dependence of the pseudocritical temperature on angular velocity and baryon chemical potential. We further examine the spatial dependence of the pseudocritical temperature in the rotating medium. Our main findings are summarized as follows: 

\begin{itemize}

    \item We find that the rotation significantly modifies the chiral condensate. The pseudocritical temperature systematically decreases with increasing angular velocity, indicating that rotation suppresses the chiral crossover temperature.
    
    \item Following the standard parametrization of the chiral transition at finite baryon chemical potential, we parameterize the rotational dependence of the pseudocritical temperature in the $T$ - $\omega$ plane. The corresponding rotational curvature coefficients are found to be $k_2^\omega=10.578\pm0.118$ and $k_4^\omega=-68.145\pm6.416$.

    \item The combined effects of rotation and finite baryon chemical potential lead to a stronger suppression of the pseudocritical temperature, demonstrating their interplay in modifying the chiral crossover. 
    
    \item Finally, we investigate the radial dependence of the pseudocritical temperature in the rotating medium and find that the suppression is strongest at larger radial distances and higher angular velocities.

\end{itemize}
The present study provides a first systematic HRG-based investigation of the rotational dependence of the chiral crossover, establishing a framework for exploring the chiral structure of rotating QCD matter in the $T$--$\mu_B$--$\omega$ space, including its spatial dependence.

\section*{Acknowledgement}
N.P. acknowledges the Ministry of Education (MoE), Government of India for the financial support. 
K.K.P. and R.S. gratefully acknowledge the DAE-DST, Govt. of India funding under the mega-science project -- “Indian participation in the ALICE experiment at CERN" bearing Project No. SR/MF/PS-02/2021-IITI (E-37123).

\end{document}